\documentclass[runningheads]{llncs}
\usepackage{graphicx}
\usepackage{amsmath}
\usepackage{hyperref}
\usepackage{cleveref}
\usepackage{csquotes}
\usepackage{xcolor}
\usepackage{comment}
\usepackage{booktabs}
\usepackage{orcidlink}
\usepackage[most]{tcolorbox}

\usepackage{glossaries}
\glsdisablehyper 
\newacronym[plural={DSLs}]{dsl}{DSL}{Domain Specific Language}
\newacronym{gui}{GUI}{Graphical User Interface}
\newacronym{mvvm}{MVVM}{Model-View-ViewModel}
\newacronym{mvc}{MVC}{Model-View-Controller}
\newacronym{mvp}{MVP}{Model-View-Presenter}
\newacronym{mlr}{MLR}{Multivocal Literature Review}
\newacronym{di}{DI}{Dependency Injection}
\newacronym{rmvrvm}{RMVRVM}{Remote-Model View Remote-View-Model}
\newacronym{mvvmc}{MVVM-C}{MVVM-Coordinator}
\newacronym{wpf}{WPF}{Windows Presentation Foundation}

\usepackage{listings,mdframed}
\usepackage{cite}
\usepackage{csquotes}

\begin{document}

\newcommand{\aspect}[2]{
\noindent
\textit{#1}: #2
}
\newcommand{\aspectItalic}[2]{
\noindent
\textit{#1}: #2
}

\newcommand{\aspectWithIndent}[2]{
\noindent
\hfill\begin{minipage}{\dimexpr\linewidth-0.4cm}
\textit{#1}: #2
\end{minipage}
}

\newcommand{\aspectWithHeadline}[2]{
\vspace{0.2cm}
\noindent
\textbf{#1}
#2
}

\newcommand{\aspectBox}[3]{
\noindent
\fbox{\begin{minipage}{0.96\linewidth}
\textbf{#1} #2: \textit{#3}
\end{minipage}}
}

\newcommand{\simpleBox}[1]{
\noindent
\fbox{\begin{minipage}{0.96\linewidth}
\textit{#1}
\end{minipage}}
}

\newcommand{\mlrproblem}[3]{
\noindent
\textbf{#1} - \textit{#2}: #3
\vspace{1mm}
}

\newcommand{\researchquestion}[2]{
\noindent
\fbox{\begin{minipage}{0.97\linewidth}
\textbf{#1:} \textit{#2}
\end{minipage}}
\vspace{2mm}
}

\newcommand{\roundIcon}[1]{\includegraphics[height=1.5ex,trim=0mm 1.5mm 0mm 1mm,]{figures/icons/#1.pdf}}

\definecolor{answer_background}{HTML}{F0F0F0}

\newcommand{\answerBox}[2]{
\begin{tcolorbox}[colframe=black, colback=answer_background, boxrule=0.5mm, boxsep=0mm]
\textbf{#1:}
\newline
#2
\end{tcolorbox}
}

\title{MVVM Revisited: Exploring Design Variants of the Model-View-ViewModel Pattern}
\titlerunning{Exploring Design Variants of the Model-View-ViewModel Pattern}
\author{Mario Fuksa\orcidlink{0000-0002-8210-094X} \and
Sandro Speth\orcidlink{0000-0002-9790-3702} \and
Steffen Becker\orcidlink{0000-0002-4532-1460}}
\institute{Institute of Software Engineering \\
University of Stuttgart, Universitätsstraße 38, 70569 Stuttgart, Germany
\email{[firstname.lastname]@iste.uni-stuttgart.de}}
\maketitle              

\begin{abstract}
Many enterprise software systems provide complex Graphical User Interfaces (GUIs) that need robust architectural patterns for well-structured software design.
However, popular GUI architectural patterns like Model-View-ViewModel (MVVM) often lack detailed implementation guidance, leading GUI developers to inappropriately use the pattern without a comprehensive overview of design variants and often-mentioned trade-offs.
Therefore, this paper presents an extensive review of MVVM design aspects and trade-offs, extending beyond the standard MVVM definition.
We conducted a multivocal literature review (MLR), including white and gray literature, to cover essential knowledge from blogs, published papers, and other unpublished formats like books.
Using the standard MVVM definition as a baseline, our study identifies (1)~76 additional design constructs grouped into 29 design aspects and (2)~16 additional benefits and 15 additional drawbacks.
These insights can guide enterprise application developers in implementing practical MVVM solutions and enable informed design decisions.

\end{abstract}
\section{Introduction}

\gls{gui} architectural patterns like \gls{mvc}, \gls{mvp}, or \gls{mvvm} play a central role when building robust and complex \glspl{gui} for enterprise applications.
Many developers use the \gls{mvvm} pattern, which promises high testability and helps to decouple the GUI from the business logic.
While Microsoft originally introduced the pattern for the \gls{wpf} application development, in recent years, the pattern has also gained more prominence for mobile development~\cite{mvvm_gossman_original_blogpost}.
For example, ViewModels are part of the suggested architecture for Android apps~\cite{mvvm_android_guide}, while it is also popular in iOS development~\cite{iOSArchitecturePatternsMvxInSwift2023}.
The origin of the \gls{mvvm} pattern is often defined in Martin Fowlers \textit{PresentationModel}, which describes the idea of separating the presentation state from the View in a dedicated observable data-structure and aims for a Humble View~\cite{patterns_fowler_PresentationModel2004, patterns_fowler_humble}.

However, while \gls{mvvm} is prominently used, it is a set of a few guidelines, and standard \gls{mvvm} definitions leave many design decisions open.
For instance, MVVM does not specify how to structure the GUI at the dialog level~\cite{engelschall2018_dissertation_hierarchicalUiCompArch}. 
Many developers have their interpretation of the pattern and use specific variants in their implementations.
This comes with architectural risks:
(1)~developers select certain \gls{mvvm} implementations without having an overview of which design alternatives they could consider.
(2)~\gls{mvvm} has implicit trade-offs, which developers often do not know in advance.

While significant research exists on \gls{mvc} and various \gls{gui} architectural patterns, no comprehensive literature study explores design variants and additional trade-offs regarding \gls{mvvm}.
Specifically, gray literature and books often contain essential aspects about the usage of \gls{mvvm}, which has not been covered by white literature so far.
The lack of a systematic review that integrates diverse sources leaves a critical void in the literature.
Therefore, many developers and researchers might miss a complete overview of \gls{mvvm}.

To fill this gap, this paper presents a \gls{mlr}, including a qualitative analysis of a broad amount of white and gray literature.
The MLR focuses on the conceptual level of the MVVM pattern and does not analyze specific GUI framework implementation details since, in our perspective, GUI frameworks do not implement or even enforce a specific MVVM design variant.
Therefore, we guide the \gls{mlr} by the two research questions:


\vspace{2.5pt}
\researchquestion{RQ1}{Which design variants do developers use when implementing \gls{mvvm}?}

\researchquestion{RQ2}{Which trade-offs do developers mention when applying \gls{mvvm}?}


As a result, we extracted 76 additional \textit{design constructs}, 16 additional \textit{benefits}, and 15 additional \textit{drawbacks}, which go beyond the \gls{mvvm} standard definition.
We synthesized 29 \textit{design aspects} to categorize those design constructs.
Therefore, this paper gives an overview of MVVM design variants and trade-offs to help developers make informed decisions when implementing MVVM.

The paper's remainder is structured as follows:
\Cref{sec:foundations} describes the \gls{mvvm} standard definition and trade-offs.
\Cref{sec:methodology} outlines the \gls{mlr} process.
\Cref{sec:discussion} discusses the results.
\Cref{sec:design_aspects} details design variants.
\Cref{sec:threats} handles threats to validity.
\Cref{sec:related_work} covers related work.
\Cref{sec:conclusion} concludes the paper.

\section{Standard Definition of Model-View-ViewModel}
\label{sec:foundations}

A central element in our \gls{mlr} is a \textit{standard definition} about \gls{mvvm}, which we use as the baseline to identify design deviations, extensions, or additional trade-offs.
Our standard definition relies mainly on the definition and trade-offs that John Gossman originally introduced in Microsoft blog posts~\cite{mvvm_gossman_original_blogpost,mvvm_gossman_original_blogpost_advantages_disadvantages}.
Additionally, we regard an often cited definition of Josh Smith on a Microsoft blog post and two further official documentation sites of Microsoft about \gls{mvvm}~\cite{mvvm_smith2009patterns,mvvm_microsoftMvvmPattern2012,mvvm_microsoftMvvmMaui2022}.

\gls{mvvm} is a GUI architectural pattern derived from \gls{mvc}, where the \textit{ViewModel} replaces the controller and uses a general data-binding mechanism.
It specializes Fowler's \textit{PresentationModel}~\cite{patterns_fowler_PresentationModel2004}.
\Cref{fig:ch2_mvvm} shows the three components:

\begin{figure}[t]
  \centering
  \vspace{-4mm}
  \includegraphics[width=0.75\linewidth]{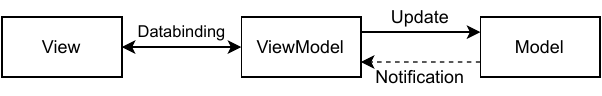}
  \vspace{-4mm}
  \caption{The Standard Model-View-ViewModel Architectural Pattern.}
  \label{fig:ch2_mvvm}
  \vspace{-5mm}
\end{figure}

\aspect{Model}{
As in \gls{mvc}, the \textit{Model} contains the data and business logic completely independent of the \gls{gui}.
The concrete design of Model classes has almost nothing to do specifically with the \gls{mvvm} pattern.
}

\aspect{View}{
Also similar to \gls{mvc}, the \textit{View} consists of visual elements (like buttons, windows, or graphics) and uses one-way\footnote{E.g., updates on a ViewModel's field are automatically reflected to a View's textbox} or two-way\footnote{E.g., in addition to one-way binding, modifications on the View's textbox are automatically reflected to the ViewModel's field} data-binding to ViewModel fields to obtain information to visualize.
The View can be data-bound directly to Model elements or by further elements defined by the ViewModels.
}

\aspect{ViewModel}{
The ViewModel handles the \textit{presentation logic} like data transformation, acts as the ``Model for the View'', and provides information by data-binding.
It exposes Commands that the View can use to interact with the Model.
ViewModels might contain (sole or extending) validation logic.
Further, the \textit{ViewModelLocator} pattern helps to instantiate and locate ViewModel instances.
}

\aspect{Relationships}{
In \gls{mvvm}, the View knows the ViewModel, and the ViewModel knows the Model.
The Model is unaware of the View and the ViewModel, while the ViewModel is unaware of the View.
}

\textit{Standard benefits} of \gls{mvvm} are that ViewModels provide an abstraction of the View and an easier way to unit-test presentation logic.
The components (View, ViewModel, Model) are decoupled from each other, supporting developers to swap, create, or maintain more easily.
It can reduce boilerplate code in the View while providing good data-binding performance.
Further, a developer-designer workflow helps the development team create robust ViewModels, while a design team can focus on user-friendly View designs.
Additionally, it cleanly separates the application's business logic and presentation logic.

\textit{Standard drawbacks} of \gls{mvvm} are the complexity for simple \glspl{gui}, challenges in designing ViewModels up-front in bigger cases, harder debugging of declarative data bindings, and increased memory consumption by binding overhead.

\section{Methodology}
\label{sec:methodology}

\begin{figure}[tb]
  \centering
  \includegraphics[width=0.999\linewidth]{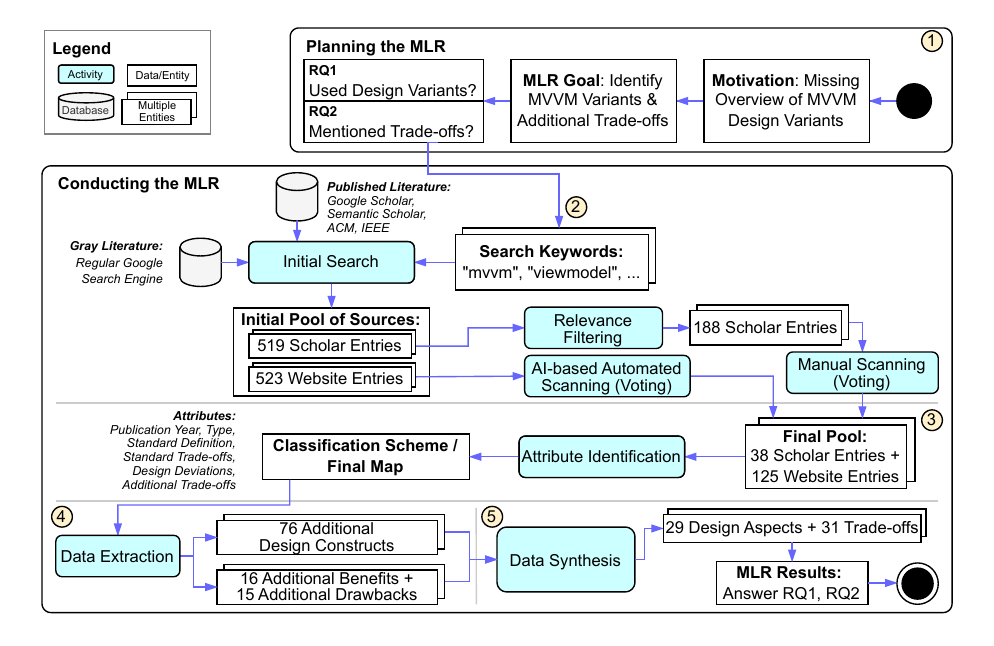}
  \caption{Overview of the Applied Multivocal Literature Review (based on Garousi~\cite{mvlr_garousi2019}).}
  \label{fig:mlr_overview}
\end{figure}

This section describes the applied \gls{mlr} process in which we conducted a qualitative analysis of \gls{mvvm}-focused sources.
We used the \gls{mlr} process of Garousi et al.~\cite{mvlr_garousi2019} to follow a structured process to review gray and published literature and extract information to answer our research questions.
\Cref{fig:mlr_overview} shows an overview of our specific process involving data entities and activities.
We separated into the planning of the \gls{mlr}, a search process, an attribute/classification design, a data extraction process, and a data synthesis.
We provide a replication package\footnote{\url{https://doi.org/10.5281/zenodo.13350488}} which transparently shows the results of each step, like the initial search, the attribute scheme with all identified design constructs and trade-offs, or the final data synthesis results.
We also included scripts to semi-automate most steps, e.g., to check for duplicate or unrelated search results.

\aspect{\roundIcon{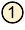} Planning the MLR}{
First, we planned the \gls{mlr} and included our motivation for the missing overview of \gls{mvvm} design variants that people apply in practice.
The goal is to identify essential variants of \gls{mvvm}, which might cover aspects not mentioned in the standard definition.
The outcomes of the planning phase are the two research questions that guide our \gls{mlr}.
}

\aspect{\roundIcon{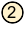} Search Process}{
The search process covers the initial search for gray and published literature, filtering, and voting.
First, we defined relevant keywords: ``mvvm'', ``model-view-viewmodel'', ``viewmodel'', and ``view model''.
We varied the combination of keywords depending on the possible search options of the databases.
We used the regular Google search engine to find gray literature and multiple databases to find published literature, such as white papers and books.
We utilized the tool \textit{Publish or Perish}\footnote{\url{https://harzing.com/resources/publish-or-perish}} to search in Google Scholar and Semantic Scholar mainly by title keywords.
Further, we did a dedicated search in the ACM and IEEE databases to complement relevant white papers that do not directly contain the keywords' titles.
The searches were up-to-date until the beginning of 2024, resulting in 519 scholar entries and 523 website entries.

\begin{table}[tb]
\caption{Exclusion Criteria of Scholar Entries.}
\label{tab:exclusion_criteria_scholar_entries}
\centering
\begin{tabular}{ll}
\toprule
Criteria        & Notes \\
\cmidrule(r){1-1}\cmidrule(r){2-2}
Not-English \hspace{1mm}     &  exclude if not written in English\\
Duplicates      &  exclude any duplicate entry\\
Unfocused       &  exclude if not focusing on \gls{mvvm} definitions\\
\bottomrule
\end{tabular}
\vspace{-5mm}
\end{table}

Next, we filtered and voted on the entries.
We first focused on scholar entries and filtered out several entries by exclusion criteria shown in \Cref{tab:exclusion_criteria_scholar_entries}, which reduced the number of scholar entries to 188.
Since scholar entries allowed us to scan efficiently based on titles, abstracts, and their typical scientific structure, we manually voted them for relevance.
Here, we also scanned \gls{mvvm} definitions if they contain no definition, only standard definition constructs, or if they potentially describe significant design constructs or additional trade-offs.
For example, we reject papers that only use \gls{mvvm} as an implementation detail without a clear \gls{mvvm} definition.
This voting resulted in 38 final scholar entries.

We used another voting approach for websites since we cannot easily filter them.
Leveraging the capabilities of ChatGPT (using GPT-4), we used AI-based automated voting.
The motivation derives from the lack of standardized website structures, which makes it difficult to scan and filter non-relevant entries without reading each website completely.
First, we manually read 20 pivot websites and classified them.
We then iteratively improved a prompt for ChatGPT, including our criteria, default definition, and trade-offs, until the pivot websites were classified as expected.
We finally used chunks of five URLs and let ChatGPT process the voting.
As a result, the AI classified the number of websites into the categories ``Standard Definition'', ``Extended Definition'', ``Extended Trade-offs'', or ``No Definition''.
The outcome of this voting is 125 website entries.

In our \gls{mlr}, we focused on the View/ViewModel-specific aspects of \gls{mvvm}.
We largely filtered out design constructs for the Model layer since they are usually not \gls{mvvm}-specific, i.e., they also apply to \gls{mvc} and \gls{mvp}.
}

\aspect{\roundIcon{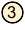} Attribute/Classification Design}{
We identified relevant attributes as a classification scheme based on the final pool of 38 scholar entries and 125 websites containing potentially significant data.
As meta-data, we are interested in the publication year and type (i.e., personal or professional articles, forum discussions, white papers, technical reports, or books).
Qualitatively, we are interested in the attribute if a source aligns with the \gls{mvvm} standard definition or standard benefits/drawbacks.
Besides this, we also classified design extensions and additional trade-offs, which extend the standard constructs.
To structure results, we then developed a helper language using JetBrains MPS (see our replication package), which prepares the structure of the classification scheme.
}

\aspect{\roundIcon{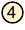} Data Extraction}{
In the data extraction phase, we analyzed the sources' data to identify relevant design constructs and trade-offs that align with our classification scheme.
This qualitative analysis carefully reviewed each voted entry using the prepared classification scheme.
Not every entry contained relevant design constructs; e.g., several websites voted by ChatGPT aligned more or less with the baseline standard definition.
%

\Cref{fig:ch3_source_topics_statistics} overviews the found design constructs and their occurrences across gray and white literature types.
We combine similar \textit{design constructs} into eleven \textit{topics} (e.g., formatting and localization) to consume the figure more easily.
The overview shows that most constructs are covered by books, followed by website articles.
Further, it highlights that the most referenced topics are navigation (e.g., how one ViewModel navigates to another View/ViewModel) and various ViewModel abstraction constructs (e.g., humble View vs. reusable ViewModels).
Therefore, we describe those two topics in Subsections \ref{sec:viewmodel_abstractions} and \ref{sec:mvvm_with_navigation} in more detail and briefly examine the further topics in the joint Subsection \ref{sec:further_design_aspects}.
Finally, we extracted 76 additional design constructs, 16 additional benefits, and 15 additional drawbacks compared to the \gls{mvvm} standard definition.
We discuss a subset of the extracted data in more detail in \Cref{sec:discussion}.
The replication package provides the full overview, including details on their occurrences and explanations.


\begin{figure}[tb]
  \centering
  \includegraphics[width=0.99\linewidth]{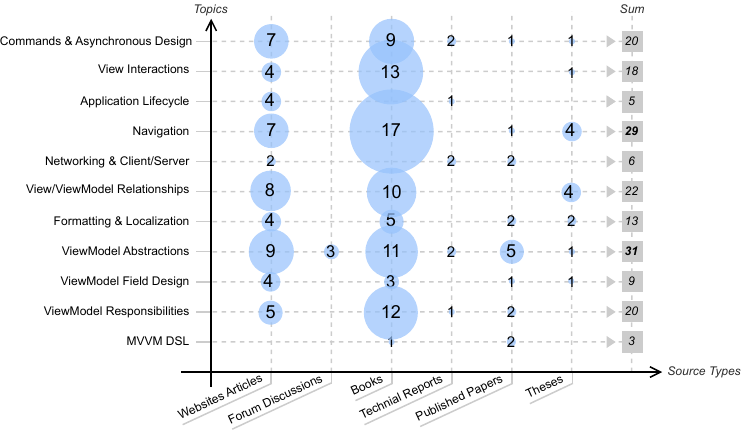}
  \caption{Distribution of 76 Design Constructs Grouped into Topics.}
  \label{fig:ch3_source_topics_statistics}
\end{figure}
}

\aspect{\roundIcon{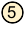} Data Synthesis}{
We performed a data synthesis from the extracted data to answer RQ1 and RQ2.
For RQ1, we classified the 76 design constructs (e.g., \textit{View has Many ViewModels}) into 29 design aspects (e.g., \textit{View/ViewModel Relationships}) and selected aspects of our particular interest to describe in \Cref{sec:design_aspects}.
For RQ2, we processed 16 additional benefits and 15 drawbacks.
In the next section, we explicitly answer the two research questions as part of the data synthesis, including a discussion of the results.
}

\section{Discussion and Results}
\label{sec:discussion}

This section discusses the results of the MVVM MLR by addressing the two research questions \textit{RQ1} (used MVVM design variants) and \textit{RQ2} (mentioned MVVM trade-offs).
We highlight key findings and practical takeaways for enterprise application developers.

\begin{figure}[b]
  \centering
  \includegraphics[width=0.99\linewidth]{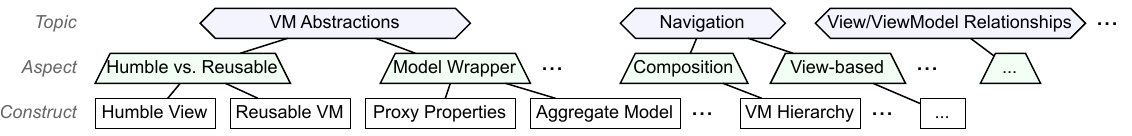}
  \vspace{-2mm}
  \caption{Design Constructs Hierarchy: Topic $\supset$ Aspect $\supset$ Construct.}
  \label{fig:ch4_construct_hierarchy}
\end{figure}

\aspectWithHeadline{MVVM Design Variants (RQ1)}{
The MLR identified 76 additional design constructs grouped into 29 design aspects, representing variants of the standard MVVM definitions.
Due to the breadth of design constructs, we further organize the 29 design aspects into eleven topics, as illustrated in \Cref{fig:ch4_construct_hierarchy}.
We focus on the two most referenced:
The \textit{ViewModel abstractions} topic includes twelve constructs in five design aspects (\textit{application structure}, \textit{coupling}, \textit{design}, \textit{humble/reusable}, \textit{model wrapper}), mentioned 31 times (see Subsection \ref{sec:viewmodel_abstractions}).
The \textit{navigation} topic includes eight constructs in three design aspects (\textit{composition}, \textit{responsibility}, \textit{view-based}) mentioned 29 times (see Subsection \ref{sec:mvvm_with_navigation}).
Further topics include command design, view interactions, lifecycle aspects, networking, View/ViewModel relationships, formatting/localization, ViewModel field design and responsibilities, or using an MVVM \gls{dsl} in Subsection \ref{sec:further_design_aspects}.
We also synthesized relations between design constructs and standard MVVM:

\begin{itemize}
    \item \textit{Restricting Constructs}: Nine constructs restrict design rules addressed by the MVVM standard definition.
    For example, while the standard definition does not limit the cardinalities between View and ViewModel, the construct \textit{View has One ViewModel} does.
    \item \textit{Extending Constructs}: 43 constructs extend the MVVM standard definition by addressing unmentioned aspects.
    For instance, \textit{Model-View-Presenter-ViewModel} and \textit{MVVM/Controller} handle the modularization of the ViewModel, addressing the often-mentioned drawback of the ViewModel growth due to many responsibilities without proper modularization.
    \item \textit{Implementing Constructs}: Twelve constructs provide concrete implementations for standard MVVM aspects.
    For example, standard definitions mention \enquote{asynchronous operations} but lack guidance for handling asynchronous data bindings not firing on the \textit{GUI thread}.
    Constructs like \textit{Asynchronous Results Handling by Mediator} address this using the mediator pattern.
    \item \textit{Confirming Constructs}: Seven constructs confirm and clarify standard MVVM \enquote{tips} or intentions by providing concrete examples.
    For instance, the \textit{Coloring in ViewModel} construct confirms the responsibility of data conversion in the ViewModel for color formatting.
    \item \textit{Divergent Constructs}: Four constructs contradict MVVM standard \enquote{tips} or intentions.
    For example, \textit{Coloring in View} contradicts the intention of placing data conversion logic into the ViewModel by describing how the View can implement the responsibility of formatting colors instead.
\end{itemize}
}

\noindent 
These design constructs provide a valuable framework for implementing the MVVM pattern, enabling developers to make well-informed design decisions.

\answerBox{RQ1 Key Takeaways}{
    The MVVM standard definitions stay vague on crucial design aspects, significantly impacting implementations.
    We identified design constructs that restrict MVVM rules for specialized design variants, address aspects not covered by standard MVVM, provide concrete implementation guidelines, and confirm or diverge from standard MVVM intentions.
}

\aspectWithHeadline{MVVM Trade-offs (RQ2)}{
    The MLR identified 16 additional \gls{mvvm} benefits and 15 additional \gls{mvvm} drawbacks (\Cref{tab:benefits_drawbacks_mvvm}), which the standard MVVM definition does not mention.
    We highlight the three most cited benefits and drawbacks here.

\noindent
    \aspect{Benefits}{First, eight sources mention the benefit that \gls{mvvm} supports easier reuse of components like the ViewModel~\cite{BuldingEnterpriseAppsWpfMvvm}.
    This is especially beneficial if a ViewModel can be reused for multiple Views.
    Second, it is stated in five sources (including multiple empirical studies) on mobile applications that \gls{mvvm} can lead to a better application performance~\cite{Wisnuadhi2020}.
    Third, four sources mention that \gls{mvvm} achieves a higher decoupling of View and ViewModel since the ViewModel usually does not know the View.
    In the so-called ``Pure MVVM'', the decoupling is further increased since the View obtains a ViewModel instance without knowing its concrete type, and data binds to its fields dynamically~\cite{MvvmSurvivalGuideSiddiqi2012}.
}
\noindent
    \aspect{Drawbacks}{Twelve sources state that the ViewModel usually has too many responsibilities.
    Inexperienced developers, in particular, might place too many responsibilities into the ViewModel without considering modularization.
    The unclear definition of \gls{mvvm} could be a possible reason~\cite{BuldingEnterpriseAppsWpfMvvm}.
    Second, seven sources discuss the high learning curve, which can hinder developers from applying the \gls{mvvm} pattern efficiently~\cite{MvvmSurvivalGuideSiddiqi2012, Anderson2012MvvmPattern, iOSArchitecturePatternsMvxInSwift2023}.
    Third, seven sources mention that \gls{mvvm} involves substantial boilerplate code, mainly if weak tooling support is used and glue code for data-binding has to be written manually~\cite{MvvmSurvivalGuideSiddiqi2012}.
}
}

\answerBox{RQ2 Key Takeaways}{
    Applying the MVVM pattern yields many benefits and drawbacks that developers might not know explicitly.
    Understanding additional trade-offs can help evaluate the pattern or its design variants more effectively and support making informed decisions when implementing MVVM.
}

\noindent
These results highlight the importance of understanding the various design constructs and trade-offs associated with the MVVM pattern.
By leveraging the additional insights from our MLR, enterprise application developers can make informed decisions and tailor their implementations to suit specific project needs better while avoiding common pitfalls.

\begin{table}[tb]
\caption{Additional Benefits and Drawbacks of MVVM (with Occurrences Number).}
\vspace{-2mm}
\label{tab:benefits_drawbacks_mvvm}
\centering
\begin{tabular}{p{0.42\textwidth}r @{\hskip 13pt} p{0.42\textwidth}r}
\toprule
\textbf{Benefit} & \textbf{No.} & \textbf{Drawback} & \textbf{No.} \\
\cmidrule(r){1-2}\cmidrule(r){3-4}
Easier Reuse of Components & 8 & Many Responsibilities in ViewModel & 12 \\
Better Performance vs. MVC/MVP & 5 & High Learning Curve & 7 \\
Increased Decoupling & 4 & Lot of Boilerplate & 7 \\
Less Boilerplate by Library & 3 & Difficult Testability & 3 \\
UI Requirements Quickly Adapted & 3 & Developer-Designer Workflow Issues & 2 \\
View Easily Replaced/Extended & 3 & Lack of Pattern Guidance & 2 \\
N-Tier: Incr. Security/Performance & 2 & Async Fetching/Threading & 2 \\
Different UI Technologies & 2 & Poor Reusability & 1 \\
Development Speed Increased & 2 & More Classes/Components & 1 \\
Easier to Cache View-state & 2 & Repeated Code in ViewModels & 1 \\
Easier Debugging & 1 & UI-Framework Features Testability & 1 \\
Less Imperative Code & 1 & Complex User Interactions Impl. & 1 \\
Well-organized Design & 1 & 3rd-Party Library Issues & 1 \\
Reduced Energy Consumption & 1 & Command Impl. Overhead & 1 \\
Reduced CPU Usage & 1 & Higher CPU Consumption & 1 \\
Easier to Maintain Lifecycle & 1 & & \\
\bottomrule
\end{tabular}
\vspace{-3mm}
\end{table}

\section{Design Aspects}
\label{sec:design_aspects}

This section discusses the resulting design aspects of the MLR.
We selected design constructs of particular interest, which we discuss in a bit more detail.

\subsection{ViewModel Abstractions}
\label{sec:viewmodel_abstractions}

This subsection discusses design constructs that focus on designing ViewModel abstractions.
The varying ViewModel abstractions significantly impact the implementation of the \gls{mvvm}.
Therefore, we discuss them in more detail.

\aspectWithHeadline{Reusable ViewModel vs. Humble View}{
There are two alternatives on how strictly a ViewModel is oriented to a specific View.
The first alternative focuses on flexibility and reusability across different Views (i.e., different information formats).
The second alternative defines a ViewModel supporting a Humble View to maximize testability and GUI framework exchangeability.
\Cref{fig:ch2_viewmodel_kinds} illustrates the distinction with a simple example of a \texttt{PersonViewModel}: a reusable ViewModel vs. a Humble View design.
Both alternatives have different impacts on reusability and testability.

\begin{figure}[th]
  \centering
  \includegraphics[width=0.95\linewidth]{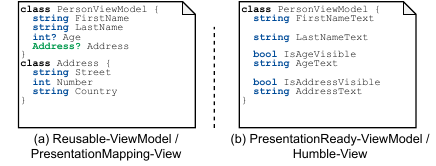}
  \vspace{-2mm}
  \caption{Reusable ViewModel vs. Humble View.}
  \vspace{-2mm}
  \label{fig:ch2_viewmodel_kinds}
\end{figure}

\aspect{Reusable-ViewModel/PresentationMapping-View}{
The first alternative defines reusable ViewModels, which can be used in multiple Views in a many-to-one relationship (\textit{ViewModel 1:n View}).
The ViewModel knows as little as possible about the specific View details and provides abstract data the View can consume.
This implies that the presentation mapping of the ViewModel data to certain GUI widget features (e.g., textbox visibility) is the responsibility of each View.
Therefore, unit testing reusable ViewModels does not cover presentation mapping logic placed in the View~\cite{FreeCodeCampMvvmAndroidPro}.
To fully cover the full presentation logic, the View also needs to be tested.
For example, a \texttt{PersonViewModel} provides more generic information like age information or an optional Address object, which the View needs to map to boolean or string representations.
}

\aspect{PresentationReady-ViewModel/Humble-View}{
The second alternative defines View-specific ViewModels, designed in a one-to-one relationship with a View (\textit{ViewModel 1:1 View}).
Unlike the first alternative, the ViewModel contains the presentation mapping logic, making it \textit{presentation-ready} with a concrete intent on how information maps to GUI widget features.
Consequently, the ViewModel provides information primarily as formatted strings or booleans, transforming the View into a \textit{Humble Object} with minimal presentation logic.
This supports unit testing of ViewModels covering most of the presentation logic, including mapping logic~\cite{FreeCodeCampMvvmAndroidPro, MvvmSurvivalGuideSiddiqi2012}.
For example, the \texttt{PersonViewModel} provides presentation-ready fields like a boolean to control the address information visibility.
Further, ViewModel fields like \texttt{FirstNameText} or \texttt{AgeText} have a concrete intent on how they should be mapped to a GUI widget.
However, a Humble View limits the ViewModel reusability across Views with different information formats.
At the same time, some sources explicitly state that reusing ViewModels should not be a premature goal~\cite{BeginningWin8AppXamlMvvm} and almost never happens in practice~\cite{LinkedInMvvmTrend}.
}
}

\aspectWithHeadline{Coupling and Model Wrappers}{
Another particularly interesting aspect from our perspective is the coupling of the ViewModel to \gls{gui} frameworks.
Suppose developers use \gls{gui} framework-specific helper classes like observables, command base classes, or visibility enumerations. In that case, the ViewModels are coupled to the framework and cannot be easily reused for other \gls{gui} frameworks in the future.
Alternatively, if developers strictly avoid using such utility classes, they might develop them themselves~\cite{MishraMvvm2017,vimotest_applicability}.
This makes the ViewModels truly independent of \gls{gui} frameworks, and the \gls{gui} framework can be migrated without touching them.
A further essential aspect of ViewModels is whether it exposes Model objects (e.g., business entities).
The reviewed sources state two options:

\aspect{Aggregate Model}{
The ViewModel directly exposes Model objects, which implies that the Model objects support observability for data-binding~\cite{MvvmSurvivalGuideSiddiqi2012}.
}

\aspect{Model Wrappers}{
Instead of exposing Model objects, the ViewModel acts as a Model wrapper and provides proxy properties of any Model property~\cite{Anderson2012MvvmPattern}.
For example, a Model \texttt{Person} class with a name is wrapped into a \texttt{PersonViewModel} with a dedicated observable name proxy property.
}
}

\subsection{MVVM with Navigation}
\label{sec:mvvm_with_navigation}

Navigation and routing of Views are important responsibilities in many enterprise applications.
This subsection discusses three design constructs: MVVM-C, hierarchical ViewModels, and 
ViewModel navigation events.

\begin{figure}[tb]
  \centering
  \includegraphics[width=0.74\linewidth]{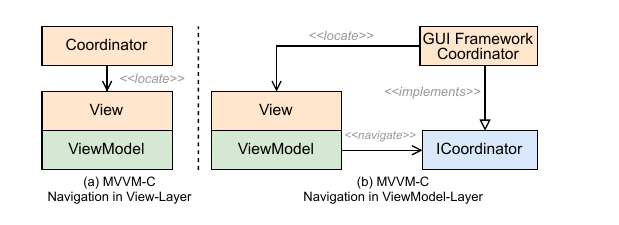}
  \vspace{-3mm}
  \caption{MVVM-C with Different Navigation Placement.}
  \label{fig:ch6_mvvm_c}
\end{figure}

\aspectWithHeadline{MVVM-C}{
    In \gls{mvvmc}, navigation is explicitly included, which extends \gls{mvvm} by a \textit{Coordinator} component responsible for navigation.
    We see two options, as illustrated in \Cref{fig:ch6_mvvm_c}:
    The first option places the coordinator into the \textit{View-layer}, solving the navigation using \gls{gui} framework-specific tooling.
    The second option introduces an abstract coordinator or navigation system in the \textit{ViewModel-layer}, providing a \gls{gui} framework-independent API for navigating from one ViewModel to another.
    The abstract coordinator accepts either a type-information about the target ViewModel or it takes a context path (e.g., a URI) to locate the target ViewModel plus context~\cite{iOSArchitecturePatternsMvxInSwift2023}.
}

\aspectWithHeadline{Hierarchical ViewModels}{
    In projects with hierarchical Views, developers can create dedicated ViewModels for each View.
    This approach mirrors the View hierarchy in the ViewModel layer.
    For instance, in a master-detail scenario, a MasterViewModel might contain a DetailViewModel object, supporting direct context navigation to the details~\cite{MvvmSurvivalGuideSiddiqi2012}.
}

\aspectWithHeadline{ViewModel Navigation Events}{
    When ViewModels are decoupled and require navigation capabilities, observer or event mechanisms offer an effective solution.
    The ViewModel triggers navigation events, which the View or an external component listens for handling navigation logic~\cite{BeginningWin8AppXamlMvvm}.
}

\subsection{Further Design Aspects}
\label{sec:further_design_aspects}

This section briefly outlines nine further topics from the \gls{mlr} results, highlighting a subset of the most relevant design constructs.
Our replication package\footnote{\url{https://doi.org/10.5281/zenodo.13350488}} discusses all design constructs in more detail and examples.

\aspectWithHeadline{Command Design and Handling of Asynchronous Results}{
\gls{wpf} introduces first-class framework support for \gls{mvvm} commands.
However, some \gls{gui} frameworks do not have such support, and developers must design commands more explicitly.
One idea is that the ViewModel provides usual methods, which are called by event handlers (e.g., \texttt{OnButtonClicked()}) in the View~\cite{Anderson2012MvvmPattern}.

Several sources mention asynchronous processing in the ViewModel (e.g., network calls on another thread).
The result handling code then has to update the ViewModel, which is usually data-bound to properties of GUI widgets and hence can only be updated from the \gls{gui} thread.
One idea is to introduce an abstracted dispatcher as a \textit{Mediator}, which provides a \gls{gui} framework-independent API to run code on the \gls{gui} thread.
Using \gls{di}, the actual \gls{gui} framework-dependent implementation is passed to the ViewModel objects, such that it can be used in result handles of asynchronous calls~\cite{Hall2010ProWpf}.

ViewModels also might prevent further actions while an asynchronous call is still running.
Developers can use a \textit{Busy Flag} to visualize information, which is set until the result handler is processed~\cite{Anderson2012MvvmPattern}.
}

\aspectWithHeadline{View Interactions}{
ViewModels often have to fulfill the requirement to interact with the View, e.g., to let the View show a message box to the user.
While \gls{mvvm}, by default, only defines that the View knows the ViewModel instance, the ViewModel cannot directly call the View.
We reviewed several design constructs to solve this problem:
(1) Introduce a View interface, similar to \gls{mvp}, which the ViewModel uses for View interaction~\cite{MishraMvvm2017}.
(2) Provide events\footnote{For example, the C\# language \texttt{event} keyword} in the ViewModel which the View can subscribe to~\cite{MvvmSurvivalGuideSiddiqi2012}.
(3) Using an interaction service that the ViewModel uses through an interface~\cite{MvvmSurvivalGuideSiddiqi2012}.
(4) Using \textit{Pub/Sub} messaging to publish/subscribe messages.
Depending on the programming languages, those options provide a way to solve the interaction problem~\cite{MvvmSurvivalGuideSiddiqi2012}.
}

\aspectWithHeadline{Application Lifecycle Aware ViewModels}{
In mobile apps (e.g., on Android), developers must manage the application lifecycle.
For example, if a user pauses an app and resumes it later.
Whenever a state is stored in ViewModels, developers should ensure that the state is valid on a resume.

In some \gls{mvvm} frameworks like Android Jetpack or MvvmCross, explicit support is provided to make ViewModels lifecycle-aware.
The idea is that ViewModels know about the application lifecycle's creation, pausing, or resuming events to control a consistent state.
A bundle object can store and restore the state, which encapsulates the persistence of data~\cite{mvvmCrossDocuViewModelLifeCycle, android_livedata}.
}

\aspectWithHeadline{Networking and Client Server}{
In client/server architectures, \gls{mvvm} can be essential in structuring the data sent over the network.
One design construct defines \textit{Remote ViewModels}, which Singh introduces in a paper as the \gls{rmvrvm} pattern~\cite{SinghRmvrvm2022}.
In \gls{rmvrvm}, the ViewModel is sent over the network while the server stores the View state to optimize further updates (i.e., send only deltas of an updated ViewModel).
Singh also discusses \gls{rmvrvm} in the context of energy efficiency~\cite{SinghRmvrvm2022}.
}

\aspectWithHeadline{View/ViewModel Relationships}{
In this aspect, we consider any statements about the View/ViewModel cardinalities that are not stated in the standard definition.
Reviewed sources mention different combinations, namely that the View has one or many ViewModels~\cite{FreeCodeCampMvvmAndroidPro, mvvm_enhanced_cross_thesis_rock_2015, mvvm_kouraklis2016} or that the ViewModel has one or many Views (e.g., when developing a wizard)~\cite{Anderson2012MvvmPattern, mvvm_kouraklis2016, FreeCodeCampMvvmAndroidPro}.
Further, some sources explicitly mention a one-to-one relationship, which implies a more strict \gls{mvvm} version.
It implies that the View and ViewModel are a \textit{tandem} developed together~\cite{LinkedInMvvmTrend, DevelopersGuidePrism2011, BeginningWin8AppXamlMvvm}.
}

\aspectWithHeadline{Formatting and Localization}{
Some sources mention design constructs on how the ViewModel or View formats data.
For example, coloring can be solved in two ways:
(1) The ViewModel provides the color of a text box (as a string color code or logical name).
(2) The ViewModel provides a logical enumeration state, and the View is responsible for mapping it to a concrete color~\cite{mvvm_kouraklis2016, DevelopersGuidePrism2011}.

Another design construct is about how the ViewModel exposes numeric information.
The ViewModel can either provide the integer or format it to the presentation-ready string, which the View directly displays to the user~\cite{mvvm_kouraklis2016}.

Multiple sources deal with the responsibility of localization.
If done in the View, the ViewModel has to provide some logical strings, which the View-layer then localizes using dictionaries.
If the ViewModel orchestrates the localization, the View is free of this responsibility, and the ViewModel uses a dictionary component that it can use to translate strings~\cite{WeissenbergModelViewDesignPatterns2019}.
}

\aspectWithHeadline{ViewModel Field Design}{
This aspect deals with how developers can design ViewModel fields concretely.
One design construct avoids Model types in ViewModels (allowed by the default definition).
It allows only using standard types like integer or string, which decouples the View/ViewModel from the Model~\cite{ManferdiniMvvmSwift}.

Another design construct focuses on visibility information in ViewModel fields.
Instead of using \gls{gui} framework-specific visibility types, ViewModels use simple boolean types\cite{Anderson2012MvvmPattern}.

Further design constructs discuss different orientations when developers design ViewModel fields:
(1) View orientation~\cite{FiveDotTwelveFlutterMvvm}.
(2) Explicitly independent of the View~\cite{Anderson2012MvvmPattern}.
(3) Model orientation by reusing Model types~\cite{Hall2010ProWpf}.
}

\aspectWithHeadline{ViewModel Responsibilities}{
When designing more complex ViewModels, developers should care about the responsibilities placed in the ViewModel.
Sources discuss different ideas, e.g., how bindings are refreshed, how dirty flags indicate state changes, or where validation occurs.

We highlight two further ideas explicitly.
First, for list items, filtering, sorting, etc., can be done in View or the ViewModel~\cite{Anderson2012MvvmPattern}.
Second, modularisation plays a key role in complex scenarios, where input logic could be placed into a separate \textit{Controller} to take this responsibility out of the ViewModel~\cite{Zarifis2017}.
}

\aspectWithHeadline{MVVM Domain-specific Languages}{
A team can leverage a \gls{dsl} to specify ViewModels and to ensure a consistent \gls{mvvm} implementation.
First, developers could use internal \gls{dsl}s by using fluent API builders, which assist in implementing ViewModel commands or data~\cite{BuldingEnterpriseAppsWpfMvvm}.
Alternatively, external \gls{dsl}s can help design a ViewModel's API programming language-independent~\cite{vimotest_applicability}.

To test ViewModels, test engineers might also utilize external \gls{dsl}s, as demonstrated by the \textit{ViMoTest} approach.
Especially when using projectional editors, GUI widgets could be pre-rendered in a test case~\cite{vimotest_applicability}.
}

\newpage
\section{Threats to Validity}
\label{sec:threats}

In this section, we discuss threats to the validity of our \gls{mlr} study.

\aspect{Construct Validity}{
We used ChatGPT to vote and filter websites automatically.
Since ChatGPT's nature is non-deterministic and sometimes unreliable, we might have included false positives and false negatives.
In particular, false negatives could negatively affect our results since we might not cover relevant aspects.
To mitigate this threat, we confirmed the correctness by checking a random selection of the voting results.

A further threat is about subjective interpretation.
The design and application of the classification scheme might involve biases or inconsistencies in categorizing and analyzing data.
Further, the manual voting process for scholar entries might introduce selection bias, affecting the relevant sources.
}

\aspect{Internal Validity}{
While we assume that our search did not scan every online resource, our initial Google search yielded over 500 websites, providing a substantial foundation.
We have not applied further methods like systematic snowballing since checking every website for references is a considerable effort.
Since we reviewed a substantial number of websites, it is unlikely that we missed crucial concepts not covered by the reviewed literature entries.
}

\aspect{External Validity}{
As professionals with specific backgrounds wrote many reviewed sources, our results might be more applicable to specific applications (e.g., enterprise applications) and less to others (e.g., mobile apps or games).

Further, many reviewed sources focus on \gls{mvvm} inherently integrated into specific technologies like \gls{wpf}.
Therefore, our findings might be limited to the ecosystems where \gls{mvvm} is commonly used.
}

\aspect{Reliability}{
The reproducibility of our search and selection process based on AI tooling like ChatGPT introduces challenges to reproduction by other researchers, impacting the reliability of the \gls{mlr} process.

Further, our data synthesizing from extracted design constructs, benefits, and drawbacks into classifications to answer research questions involves subjective judgment, which might vary among researchers.
}

\section{Related Work}
\label{sec:related_work}

This section discusses related work about \gls{mvvm} or MV* overview studies.

Wongtanuwat et al. created a systematic guideline on detecting the correctness when applying \gls{mvvm} in Objective-C programs~\cite{mvvm_wongtanuwat2020Violations}.
Weissenberg discusses best practices and lessons learned using the \gls{mvvm} pattern in an industrial WPF application~\cite{WeissenbergModelViewDesignPatterns2019}.
While these papers specifically discuss the \gls{mvvm} pattern, they are context-specific and do not analyze \gls{mvvm} in a literature review.

Lou compared the \gls{mvc}, \gls{mvp}, and \gls{mvvm} patterns for native Android app architectures regarding testability, modifiability, and performance~\cite{android_mvx_comparison_lou2016}.
Similarly, Sholichin et al. reviewed \gls{mvc}, \gls{mvp}, \gls{mvvm}, and VIPER in the context of iOS architectural patterns~\cite{sholichin2019iosPatternReview}.
Further, Magics-Verkman et al. compared \gls{mvc}, \gls{mvvm}, and MVI for testability and performance in iOS mobile application development~\cite{magicsVerkmanArchitecturalPatternsIos}.
These studies used concrete implementations of \gls{mvvm}.
They quantitatively compared them to other GUI architectural patterns for quality attributes, while our study qualitatively analyses \gls{mvvm} by a literature review.

Lappalainen and Kobayashi qualitatively compared \gls{mvc}, \gls{mvp}, and \gls{mvvm} by reviewing literature~\cite{mvvm_Lappalainen2017APL}.
Syromiatnikov and Weyns selected several GUI architectural patterns like \gls{mvvm}, reviewed sources describing those patterns, and qualitatively classified them as a landscape of GUI design patterns~\cite{Syromiatnikov_2014}.
While these studies qualitatively review \gls{mvvm}, they focus on a more extensive landscape of GUI architectural patterns and use no systematic literature survey.

Daoudi et al. empirically studied the occurrence of \gls{mvc}, \gls{mvp}, or \gls{mvvm} in Android apps~\cite{daoudiExplorativeMvcAndroid2019}.
Chekhaba et al. introduced the machine learning tool Coach to identify \gls{mvc}, \gls{mvp}, or \gls{mvvm} in Android apps~\cite{chekhabaCoach2021}.
Unlike our study, they do not qualitatively analyze design variants of \gls{mvvm} using literature reviews.

Verdecchia et al. performed a systematic mixed-method empirical study on Android app architectures, including semi-structured interviews, gray literature, and white literature~\cite{verdecchiaGuidelinesAndroidArchitecture2019}.
While they review the literature, our paper focuses specifically on \gls{mvvm} and analyses design aspects and trade-offs more deeply.

\section{Conclusion}
\label{sec:conclusion}

This paper presented an \gls{mlr} with a qualitative analysis of white and gray literature about the \gls{mvvm} pattern.
We used the standard definition and standard trade-offs of \gls{mvvm} from familiar standard sources like Gossman's original blog post, which introduced \gls{mvvm}.
Based on a selection of 519 scholar entries and 523 websites, we filtered out 38 scholar entries and 125 websites, which potentially extend design aspects or trade-offs compared to the standard definition.
We then extracted 76 additional design constructs, 16 additional benefits, and 15 additional drawbacks.
Finally, we categorized the design constructs into 29 design aspects and further grouped those aspects into eleven topics.
We briefly described a subset of design constructs in the paper, while we published a detailed replication package with the results of the planning phase, search and selection process, data extraction, and synthesis.

The synthesized design aspects and trade-offs provide an overview of design variants using the MVVM pattern.
Practitioners, such as enterprise application developers, can utilize this overview as a catalog of potential solutions when implementing MVVM and as a checklist to ensure considering common design aspects.
Researchers can also benefit from this overview using the selected sources or the results of the replication package when further studying \gls{mvvm}.

Future work could study the extracted design aspects and trade-offs, systematically analyzing conflicts between design constructs and their associated trade-offs.
Such an analysis could lead to an assessment of the design constructs and recommendations on which constructs to adopt and which to avoid.
Additionally, there is the potential to create formal pattern descriptions of both the standard MVVM definition and a subset of relevant design variants.
Moreover, while we studied literature to identify MVVM design aspects and solutions, scanning public code like GitHub repositories for MVVM implementations could reveal design constructs not covered by literature.
Furthermore, we plan to elaborate on the Humble View idea by applying the pattern within our doctoral ViMoTest project, where the ViewModel provides a presentation-ready abstraction by directly aligning with GUI widgets~\cite{vimotest_docsym, vimotest_applicability}.

\newpage
\bibliographystyle{splncs04}
\bibliography{main}

\begin{thebibliography}{10}
\providecommand{\url}[1]{\texttt{#1}}
\providecommand{\urlprefix}{URL }
\providecommand{\doi}[1]{https://doi.org/#1}

\bibitem{Anderson2012MvvmPattern}
Anderson, C.: The Model-View-ViewModel (MVVM) Design Pattern, pp. 461--499.
  Apress, Berkeley, CA (2012). \doi{10.1007/978-1-4302-3501-9\_13}

\bibitem{DevelopersGuidePrism2011}
Brumfield, B., Cox, G., Hill, D., Noyes, B., Puleio, M., Shifflett, K.:
  {Developer's Guide to Microsoft Prism 4: Building Modular MVVM Applications
  with Windows Presentation Foundation and Microsoft Silverlight}. Microsoft
  Press (2011), {ISBN: 978-0-73565-610-9}

\bibitem{BeginningWin8AppXamlMvvm}
Burns, K.: Introducing MVVM, pp. 127--140. Apress, Berkeley, CA (2012).
  \doi{10.1007/978-1-4302-4567-4\_9}

\bibitem{chekhabaCoach2021}
Chekhaba, C., Rebatchi, H., ElBoussaidi, G., Moha, N., Kpodjedo, S.: {Coach:
  classification-based architectural patterns detection in Android apps}. In:
  Proceedings of the 36th Annual ACM Symposium on Applied Computing. p.
  1429–1438. SAC '21, Association for Computing Machinery, New York, NY, USA
  (2021). \doi{10.1145/3412841.3442018}

\bibitem{daoudiExplorativeMvcAndroid2019}
Daoudi, A., ElBoussaidi, G., Moha, N., Kpodjedo, S.: {An exploratory study of
  MVC-based architectural patterns in Android apps}. In: Proceedings of the
  34th ACM/SIGAPP Symposium on Applied Computing. p. 1711–1720. SAC '19,
  Association for Computing Machinery, New York, NY, USA (2019).
  \doi{10.1145/3297280.3297447}

\bibitem{engelschall2018_dissertation_hierarchicalUiCompArch}
Engelschall, R.S.: Hierarchical user interface component architecture.
  BoD--Books on Demand (2018)

\bibitem{patterns_fowler_PresentationModel2004}
Fowler, M.: {Presentation Model} (Jul 2004),
  \url{https://martinfowler.com/eaaDev/PresentationModel.html}, accessed:
  2024-06-18

\bibitem{patterns_fowler_humble}
Fowler, M.: {HumbleObject} (Apr 2020),
  \url{https://martinfowler.com/bliki/HumbleObject.html}, accessed: 2024-06-18

\bibitem{vimotest_docsym}
Fuksa, M.: {ViMoTest: A Low Code Approach to Specify ViewModel-Based Tests with
  a Projectional DSL Using JetBrains MPS}. In: Proceedings of the 25th
  International Conference on Model Driven Engineering Languages and Systems:
  Companion Proceedings. p. 189–194. MODELS '22, Association for Computing
  Machinery, New York, NY, USA (2022). \doi{10.1145/3550356.3558513}

\bibitem{vimotest_applicability}
Fuksa, M., Speth, S., Becker, S.: {Applicability of the ViMoTest Approach for
  Automated GUI Testing: A Field Study}. In: 2023 ACM/IEEE International
  Conference on Model Driven Engineering Languages and Systems Companion
  (MODELS-C). pp. 821--830 (2023). \doi{10.1109/MODELS-C59198.2023.00131}

\bibitem{iOSArchitecturePatternsMvxInSwift2023}
Garc{\'i}a, R.F.: MVVM: Model--View--ViewModel, pp. 145--224. Apress, Berkeley,
  CA (2023). \doi{10.1007/978-1-4842-9069-9\_4}

\bibitem{BuldingEnterpriseAppsWpfMvvm}
Garofalo, R.: Building enterprise applications with Windows Presentation
  Foundation and the model view ViewModel Pattern. Microsoft Press (2011),
  {ISBN: 978-0-73565-092-3}

\bibitem{mvlr_garousi2019}
Garousi, V., Felderer, M., Mäntylä, M.V.: Guidelines for including grey
  literature and conducting multivocal literature reviews in software
  engineering. Information and Software Technology  \textbf{106},  101--121
  (2019). \doi{10.1016/j.infsof.2018.09.006}

\bibitem{mvvm_android_guide}
Google, n.d., O.H.A.: Guide to app architecture (Feb 2021),
  \url{https://developer.android.com/jetpack/guide}, accessed: 2024-06-18

\bibitem{android_livedata}
Google, n.d., O.H.A.: {LiveData overview} (Feb 2024),
  \url{https://developer.android.com/topic/libraries/architecture/livedata},
  accessed: 2024-06-18

\bibitem{mvvm_gossman_original_blogpost}
Gossman, J.: Introduction to model/view/viewmodel pattern for building wpf apps
  (Oct 2005),
  \url{https://docs.microsoft.com/de-de/archive/blogs/johngossman/introduction-to-modelviewviewmodel-pattern-for-building-wpf-apps},
  accessed: 2024-06-18

\bibitem{mvvm_gossman_original_blogpost_advantages_disadvantages}
Gossman, J.: {Advantages and disadvantages of M-V-VM} (Apr 2006),
  \url{https://docs.microsoft.com/en-us/archive/blogs/johngossman/advantages-and-disadvantages-of-m-v-vm},
  accessed: 2024-06-18

\bibitem{Hall2010ProWpf}
Hall, G.M.: The ViewModel, pp. 81--110. Apress, Berkeley, CA (2010).
  \doi{10.1007/978-1-4302-3163-9\_4}

\bibitem{FreeCodeCampMvvmAndroidPro}
Kay, R.M.: {How to Use Model-View-ViewModel on Android Like a Pro}.
  \url{https://www.freecodecamp.org/news/model-view-viewmodel-android-tutorial}
  (Dec 2020), accessed: 2024-06-18

\bibitem{mvvm_kouraklis2016}
Kouraklis, J.: MVVM as Design Pattern, pp. 1--12. Apress, Berkeley, CA (2016).
  \doi{10.1007/978-1-4842-2214-0\_1}

\bibitem{mvvm_Lappalainen2017APL}
Lappalainen, S., Kobayashi, T.: {A Pattern Language for MVC Derivatives}. In:
  Proc. 6th Asian Conference on Pattern Languages of Programs (2017),
  \url{http://www.washi.cs.waseda.ac.jp/wp-content/uploads/2017/03/Sami-Lappalainen.pdf},
  accessed: 2024-06-18

\bibitem{android_mvx_comparison_lou2016}
Lou, T.: {A Comparison of Android Native App Architecture – MVC, MVP and
  MVVM}. Master's thesis, Aalto University. School of Science (2016),
  \url{http://urn.fi/URN:NBN:fi:aalto-201610124940}

\bibitem{magicsVerkmanArchitecturalPatternsIos}
Magics-Verkman, H., Zmaranda, D.R., Gy\H{o}r\"odi, C.A., Gy\H{o}r\"odi, R.c.:
  {A Comparison of Architectural Patterns for Testability and Performance
  Quality for iOS Mobile Applications Development}. In: 2023 17th International
  Conference on Engineering of Modern Electric Systems (EMES). pp.~1--4 (2023).
  \doi{10.1109/EMES58375.2023.10171619}

\bibitem{ManferdiniMvvmSwift}
Manferdini, M.: {MVVM in SwiftUI for a Better Architecture}.
  \url{https://matteomanferdini.com/mvvm-swiftui} (Dec 2023), accessed:
  2024-06-18

\bibitem{mvvm_microsoftMvvmPattern2012}
{Microsoft}: {The MVVM Pattern} (2012),
  \url{https://learn.microsoft.com/en-us/previous-versions/msp-n-p/hh848246(v=pandp.10)},
  accessed: 2024-06-18

\bibitem{mvvm_microsoftMvvmMaui2022}
{Microsoft}: {Model-View-ViewModel (MVVM)}.
  \url{https://learn.microsoft.com/en-us/dotnet/architecture/maui/mvvm} (2022),
  accessed: 2024-06-18

\bibitem{MishraMvvm2017}
Mishra, A.: The MVVM Architectural Pattern, pp. 43--60. Apress, Berkeley, CA
  (2017). \doi{10.1007/978-1-4842-2689-6\_3}

\bibitem{FiveDotTwelveFlutterMvvm}
Moliński, D.: {Flutter architecture: implementing the MVVM pattern} (Feb
  2022),
  \url{https://fivedottwelve.com/blog/flutter-architecture-implementing-the-mvvm-pattern},
  accessed: 2024-06-18

\bibitem{mvvmCrossDocuViewModelLifeCycle}
MvvmCross: {Introduction to Model/View/ViewModel pattern for building WPF
  apps}.
  \url{https://www.mvvmcross.com/documentation/fundamentals/viewmodel-lifecycle}
  (Aug 2023), accessed: 2024-06-18

\bibitem{mvvm_enhanced_cross_thesis_rock_2015}
Rock, V.: {Using MVVM for enhanced cross platform development of mobile and
  desktop application}. Master's thesis, Master’s Thesis (2015),
  \url{https://diglib.tugraz.at/using-mvvm-for-enhanced-cross-platform-development-of-mobile-and%2Ddesktop-applications-2015},
  accessed: 2024-06-18

\bibitem{sholichin2019iosPatternReview}
Sholichin, F., Isa, M.A.B., Halim, S.A., Harun, M.F.B.: {Review Of IOs
  Architectural Pattern For Testability, Modifiability, And Performance
  Quality}. Journal Of Theoretical And Applied Information Technology
  \textbf{97}(15) (2019),
  \url{https://www.jatit.org/volumes/Vol97No15/3Vol97No15.pdf}, accessed:
  2024-06-18

\bibitem{SinghRmvrvm2022}
Singh, L.: {RMVRVM – A Paradigm for Creating Energy Efficient User
  Applications Connected to Cloud through REST API}. In: 15th Innovations in
  Software Engineering Conference. ISEC 2022, Association for Computing
  Machinery, New York, NY, USA (2022). \doi{10.1145/3511430.3511434}

\bibitem{mvvm_smith2009patterns}
Smith, J.: {Patterns - WPF Apps With The Model-View-ViewModel Design Pattern}
  (2009),
  \url{https://learn.microsoft.com/en-us/archive/msdn-magazine/2009/february/patterns-wpf-apps-with-the-model-view-viewmodel-design-pattern},
  accessed: 2024-06-18

\bibitem{LinkedInMvvmTrend}
Stein, G.: {Introduction to Model/View/ViewModel pattern for building WPF
  apps}. \url{https://www.linkedin.com/pulse/mvvm-fashion-trend-gregory-stein}
  (Mar 2021), accessed: 2024-06-18

\bibitem{Syromiatnikov_2014}
Syromiatnikov, A., Weyns, D.: {A Journey through the Land of Model-View-Design
  Patterns}. In: 2014 IEEE/IFIP Conference on Software Architecture. pp. 21--30
  (2014). \doi{10.1109/WICSA.2014.13}

\bibitem{verdecchiaGuidelinesAndroidArchitecture2019}
Verdecchia, R., Malavolta, I., Lago, P.: Guidelines for architecting android
  apps: A mixed-method empirical study. In: 2019 IEEE International Conference
  on Software Architecture (ICSA). pp. 141--150 (2019).
  \doi{10.1109/ICSA.2019.00023}

\bibitem{MvvmSurvivalGuideSiddiqi2012}
Vice, R., Siddiqi, M.S.: MVVM Survival Guide for Enterprise Architectures in
  Silverlight and WPF. Packt Publishing Ltd (2012), {ISBN: 978-1-84968-342-5}

\bibitem{WeissenbergModelViewDesignPatterns2019}
Weissenberg, C.: {Model-View Design Patterns}. Tagungsband p.~102 (2019),
  {ISBN: 978-3-00-064236-4}

\bibitem{Wisnuadhi2020}
Wisnuadhi, B., Munawar, G., Wahyu, U.: {Performance Comparison of Native
  Android Application on MVP and MVVM}. In: Proceedings of the International
  Seminar of Science and Applied Technology (ISSAT 2020). pp. 276--282.
  Atlantis Press (2020). \doi{10.2991/aer.k.201221.047},
  \url{https://doi.org/10.2991/aer.k.201221.047}

\bibitem{mvvm_wongtanuwat2020Violations}
Wongtanuwat, W., Senivongse, T.: {Detection of Violation of MVVM Design Pattern
  in Objective-C Programs}. In: Proceedings of the 8th International Conference
  on Computer and Communications Management. p. 54–58. ICCCM '20, Association
  for Computing Machinery, New York, NY, USA (2020).
  \doi{10.1145/3411174.3411193}

\bibitem{Zarifis2017}
Zarifis, K., Papakonstantinou, Y.: {In-depth Survey of MVVM Web Application
  Frameworks}. Tech. rep., Technical report of UCSDSE, University of California
  (2016), \url{https://dbucsd.github.io/paperpdfs/2016_4.pdf}, accessed:
  2024-06-18

\end{thebibliography}

\end{document}